\documentclass{revtex4}
\usepackage{latexsym}
\usepackage{epsfig}

\usepackage{amsmath}
\usepackage{amsfonts}
\usepackage{amssymb}
\usepackage{graphicx}

\newcommand{\beq}{\begin{equation}}
\newcommand{\eeq}{\end{equation}}
\newcommand{\bea}{\begin{eqnarray}}
\newcommand{\eea}{\end{eqnarray}}

\def\p{\partial}
\def\Tdot#1{{{#1}^{\hbox{.}}}}
\def\Tddot#1{{{#1}^{\hbox{..}}}}
\def\D{{D}}
\def\d{{\delta}}
\def\T{{\bf T}}
\def\R{{\cal K}}
\def\L{{\cal L}_u}
\def\HH{H}

\def\mP{M_P}

\def\r{\rangle}
\def\R{{\cal R}}

\def\u{c}

\def\Xf{X_1}
\def\Xs{X_2}
\def\af{{\delta \alpha}}
\def\as{{\delta \alpha}^{(2)}}
\def\sif{{\delta \sigma}}
\def\sis{{\delta \sigma}}
\def\sf{{\delta s}}
\def\ss{{\delta s}}
\def\phf{{\delta \varphi}}
\def\phs{{\delta \varphi}^{(2)}}
\def\phif{{\delta \phi}}
\def\phis{{\delta \phi}}

\def\chif{{\delta \chi}}
\def\chis{{\delta \chi}^{(2)}}
\def\sif{{\delta \sigma}}
\def\rhof{{\delta \rho}}
\def\rhos{{\delta \rho}}
\def\ps{{\delta P}^{(2)}}
\def\pf{{\delta P}}
\def\ab{\bar{\alpha}}
\def\rhob{\bar{\rho}}
\def\pb{\bar{P}}
\def\sib{{\bar{\sigma}}}
\def\phib{\bar{\phi}}
\def\chib{\bar{\chi}}
\def\sb{\bar{s}}
\def\thetab{\bar{\theta}}

\def\Qs{Q^{(2)}}
\def\Qf{Q}

\def\x{{\xi}}

\begin{document}



\title{A geometrical approach to  nonlinear perturbations in relativistic cosmology}

\author{David Langlois$^{a}$, Filippo Vernizzi$^b$}
\affiliation{$^a$ APC (CNRS-Universit\'e Paris 7), 10
rue Alice Domon et L\'eonie Duquet, 75205 Paris Cedex 13, France}
\affiliation{$^b$CEA, IPhT, 91191 Gif-sur-Yvette
c\'edex,
France,
CNRS, URA-2306, 91191 Gif-sur-Yvette c\'edex, France}


\vspace{1cm}

\begin{abstract}
We give a pedagogical  review of  a covariant and fully non-perturbative approach to study nonlinear perturbations in cosmology. In the first part, devoted to  cosmological fluids,  we  define a nonlinear extension of the uniform-density curvature perturbation and derive its evolution equation. In the second part, we focus our attention on multiple scalar fields and present a  nonlinear description in terms of  adiabatic and entropy perturbations. In both cases, we show how the formalism presented here enables one to  easily obtain equations up to second, third  and higher orders.
\end{abstract}

\maketitle

\date{\today}


\section{Introduction}
The relativistic theory of cosmological perturbations 
(see e.g.  \cite{Bardeen:1980kt,Kodama:1985bj,Mukhanov:1990me,Durrer:1993db,Langlois:2004de,Malik:2008im,Langlois:2010xc})
is an indispensable tool to interpret cosmological data such as the Cosmic
Microwave Background (CMB) anisotropies, and thus to connect
 the scenarios of the early universe, such as inflation,
to  cosmological observations. Because the temperature
anisotropies of the CMB are so small ($\delta T/T \sim 10^{-5}$),
considering only {\it linear} perturbations is an excellent first
approximation and  this is  why most of the efforts devoted to the study
of cosmological perturbations have dealt with the linear
theory.

However, with the rapidly increasing precision of cosmological data, we have now reached a stage 
where nonlinear features of primordial perturbations could be observationally accessible. 
In the last few years, this has motivated  an intensive development of the theory of cosmological perturbations beyond linear order. 
Two main strategies have been considered. The first one is based on extending the traditional coordinate-based approach by adding second-order perturbations in the metric and in the matter fields and by computing the equations of motion for the metric and matter perturbations by brute force \cite{Tomita,Bruni:1996im,Matarrese:1997ay,Bartolo:2004if,Noh:2004bc,Malik:2008im}. 
The second strategy is based on a fully nonlinear treatment, but restricted to super-Hubble scales and working directly in a coordinate system \cite{Salopek:1990jq,Comer:1994np,Deruelle:1994iz,Lyth:2003im,Rigopoulos:2003ak,Kolb:2004jg,Lyth:2004gb}. This strategy is related to the so called separate
universe picture that represents our universe, on scales larger
than the Hubble radius, as juxtaposed
Friedmann-Lemaitre-Robertson-Walker (FLRW) universes with
slightly different scale factors \cite{Starobinsky:1986fxa,Sasaki:1995aw}.

The approach that we present here, based on \cite{Langlois:2005ii,Langlois:2005qp,Langlois:2006iq,Langlois:2006vv}, relies on a geometrical perspective. From a  computational point of view, it  can be seen as a ``middle way'' between the two main strategies discussed above. Inspired by the so-called covariant formalism \cite{Hawking:1966qi,Ellis} for cosmological perturbations developed in \cite{Ellis:1989jt,Ellis:1989ju,Ellis:1990gi,Bruni:1991kb,Dunsby:1991xk,Bruni:1992dg} (see also the article by R. Maartens in this issue \cite{maartens}), the fundamental idea is to use a nonlinear and covariant approach as long as possible and try to construct covariant quantities that mimic the traditional quantities that have been useful in the theory of linear perturbations (see also \cite{Clarkson:2003af} for an alternative application of the covariant formalism to nonlinear perturbations). Since these quantities are tensors, they are intuitively much easier to understand from a geometrical point of view. Once these covariant objects have been defined and their evolution equation obtained in a fully nonlinear and covariant form, it is then possible to choose a coordinate system and to expand the quantities and equations to the desired order, in order to make concrete quantitative calculations.

In this review we discuss two examples of this approach which are particularly illuminating. The first is the nonlinear extension of the familiar linear curvature perturbation on uniform-density hypersurfaces, usually denoted $\zeta$,
introduced in \cite{Bardeen:1983qw}. This quantity plays a crucial role in the linear theory, because 
it is  conserved on large scales for adiabatic perturbations \cite{Hwang,Wands:2000dp}, and therefore enables one to very easily relate cosmological perturbations in the very early universe and in the late universe, as long as the scales remain super-Hubble. The second example is useful in the context of multi-field inflation. In the linear theory, it is useful to decompose the perturbations of the scalar fields into (instantaneous) adiabatic and entropic modes, which are simply the projections of the perturbations along directions that are, respectively, tangential and orthogonal to the background trajectory in field space. Once again, it is possible to generalize this decomposition into adiabatic and entropic modes at the nonlinear level. This can be used to compute the evolution of nonlinear perturbations and the resulting non-Gaussianities. 

This review is divided into two parts. In the next section we consider the treatment of a cosmological fluid, while  the third,  and final, section is devoted  to the case of several scalar fields.

\section{Cosmological fluids} \label{sec:choices}

\subsection{Covariant approach}

In this section  we consider 
a single perfect fluid,  characterized by a comoving
four-velocity $u^a$ ($u_a u^a =-1$), a proper energy density
$\rho$ and a pressure $P$.
The energy-momentum tensor associated to the perfect fluid is given by
\beq
T^a_{\ b}=\left(\rho+P\right) u^a u_b+Pg^a_{\ b}. \label{emt}
\eeq
To fully characterize the fluid, one needs an equation of state
relating $P$ to $\rho$ and, possibly, to other physical quantities
if the fluid is not barotropic.

The spatial projection tensor orthogonal to the fluid velocity
$u^a$ is defined by
\beq
h_{ab}=g_{ab}+u_a u_b, \quad \quad (h^{a}_{\ b} h^b_{\ c}=h^a_{\
c}, \quad h_a^{\ b}u_b=0).
\eeq
It is also useful to introduce the familiar decomposition
\beq
\nabla_b u_a=\sigma_{ab}+\omega_{ab}+{1\over 3}\Theta
h_{ab}-a_a u_b, \label{decomposition}
\eeq
where one finds on the right-hand side  the (symmetric) shear tensor $\sigma_{ab}$, the
(antisymmetric) vorticity  tensor $\omega_{ab}$, the volume expansion $\Theta \equiv \nabla_a u^a$ and the acceleration $a_a\equiv u^c \nabla_c u^a$.

The integration of $\Theta$ along the fluid world lines with
respect to the associated proper time $\tau$,
\beq
\label{alpha_def} \alpha \equiv {1\over 3}\int d\tau \, \Theta,
\eeq
can be used to define,  for each observer comoving with the
fluid, a local scale factor $S=e^\alpha$. It follows that
$\Theta = 3 \dot\alpha$,
where the dot of a scalar quantity denotes its derivative along $u^a$, i.e.~$\dot{\alpha} \equiv  u^a \nabla_a \alpha$.
This quantity $\alpha$, which can be seen as a (covariantly defined) local number of e-folds, plays a crucial r\^ole in the definition of our covariant and nonlinear extension of  the familiar linear curvature perturbation on
uniform-density hypersurfaces, usually denoted $\zeta$,
introduced in \cite{Bardeen:1983qw}. 

Our starting point is the 
conservation of the energy-momentum tensor,
\beq
\label{conserv}
\nabla_a T^a_{\ b}=0,
\eeq
which yields, after substituting (\ref{emt}) and projecting along $u^b$, the continuity equation
\beq
\label{continuity} \dot\rho + 3 \dot \alpha (\rho + P)=0.
\eeq
The spacetime gradient of this expression can be written as 
\beq
\label{grad_continuity} \Tdot{(\p_a \rho)}+ 3\, \Tdot{(\p_a \alpha)}\,  (\rho + P) + 3 \dot \alpha\,  \p_a (\rho+P)=0,
\eeq
where the dot acting on a spacetime gradient, which is a covector,  is defined as the {\em Lie derivative}  (see e.g. \cite{Wald:1984rg})  with respect to $u^a$. For any covector $X_a$, this means 
\beq
\dot X_a \equiv {\cal L}_u X_a \equiv  u^b \partial_b X_a + X_b \partial_a u^b.
\label{Lie_def}
\eeq
More generally, a dot acting on any tensor will denote the Lie derivative of this tensor with respect to $u^a$. This definition is compatible with our notation for the scalars since the Lie derivative coincides with the derivative along $u^a$ for scalars (and we have used the property $\partial_a(\dot\rho)=\Tdot{(\partial_a \rho)}$ to obtain (\ref{grad_continuity})).

After some simple manipulations, one finds that Eq.~(\ref{grad_continuity}) is equivalent to \cite{Langlois:2005ii,Langlois:2005qp}
\beq
\label{dot_zeta}
\dot\zeta_a=
-\frac{\Theta}{3(\rho+p)}\left( \p_a p -
\frac{\dot p}{\dot \rho} \p_a\rho\right) \, ,
\eeq
where the left-hand side consists of the time derivative, i.e.~the Lie derivative with respect to $u^a$, of  the covector
\beq
 \zeta_a\equiv
\p_a\alpha-\frac{\dot\alpha}{\dot\rho}\p_a\rho\, .
\label{zeta_a}
\eeq
This covector    can also be re-expressed, using (\ref{continuity}), 
 in the form
\beq
\label{zeta_a2}
\zeta_a=\p_a\alpha+\frac{\p_a\rho}{3(\rho+P)} \,
\eeq
and, if $w\equiv P/\rho$ is constant, this  is a total gradient since
\beq
\zeta_a=\p_a\left[\alpha+\frac{1}{3(1+w)}\ln \rho\right]\qquad (w=P/\rho={\rm const}).
\label{zeta_a_w}
\eeq

On the right-hand side,
 the quantity
 \beq
 \label{Gamma}
 \Gamma_a\equiv
\p_aP- {\dot P\over \dot\rho}\p_a\rho
\eeq
 is a  nonlinear generalization of the
non-adiabatic pressure. It vanishes for purely adiabatic perturbations, for instance when the
pressure $P$ is solely a function of the density $\rho$. 

Eq.~(\ref{dot_zeta})  has a form very similar to the conservation equation
for $\zeta$ of the linear theory, which will be rederived in the
next section. 
At the non-linear level, our covector $\zeta_a$ is closely related to the non-linear perturbation introduced in 
\cite{Lyth:2004gb},
\beq
\zeta=\delta N +\frac13\int_{\bar{\rho}}^\rho \frac{d\tilde\rho}{\tilde\rho+\tilde P},
\eeq 
where $\bar\rho(t)$ is the homogeneous background density, while $\rho(t, \vec{x})$ is the local energy density and $N=\int dt \, \Theta/3$ is the number of e-folds, defined with respect to the time coordinate $t$. A similar non-linear quantity has been introduced earlier in  \cite{Rigopoulos:2003ak}. 
The advantage of the definition (\ref{zeta_a}) is that one does not need to introduce a coordinate system or to restrict oneself to super-Hubble scales, and that its evolution equation (\ref{dot_zeta}) is 
 {\it exact},  fully
{\it non-perturbative} and valid at {\it all scales}. Moreover, it is worth stressing that this equation is a  direct consequence of the conservation of the energy-momentum tensor and is independent of the underlying theory of gravitation.

\subsection{Link with the coordinate approach}

We now relate the covariant approach with the more familiar
coordinate based formalism \cite{Bardeen:1980kt,Kodama:1985bj,Mukhanov:1990me,Durrer:1993db,Langlois:2004de,Malik:2008im,Langlois:2010xc}. We first examine the linear
perturbations and consider later the perturbations  
at second and third  orders.

\subsubsection{Linear theory} \label{Linear}
A cosmological spacetime closed to FLRW geometry can be described by 
the perturbed metric 
\beq
ds^2=-(1+2A) dt^2+ 2 aB_i dx^i dt + a^2\left( \gamma_{ij}+  H_{ij}\right) dx^i dx^j\, ,
\eeq
where  $a=e^{{\bar\alpha}}$ is the background scale factor. As usual, one can decompose the linear perturbations 
into so-called scalar, vector and tensor modes,
\beq
\begin{split}
B_i  = & \vec \nabla_i B + B^V_i, \\
\label{H} H_{ij} = & 
-2\psi \gamma_{ij}+2 \vec\nabla_i\vec\nabla_j E+2
\vec \nabla_{(i}E^V_{j)}+2 { E}^T_{ij},
\end{split}
\eeq
where $B_i^V$ and $E^V_i$ are transverse, i.e., 
$\vec \nabla_i B^V{}^i=0=\vec \nabla_iE^V{}^i$, and 
${E}^T_{ij}$ is transverse and traceless, i.e., $\vec
\nabla_i{E}^T{}^{ij}=0$ and $\gamma^{ij}{E}^T_{ij}=0$. Here $\vec\nabla_i$
denotes the three-dimensional covariant derivative with respect 
to the homogeneous
spatial metric $\gamma_{ij}$ (which is also used to lower or raise
the spatial indices).

The corresponding matter content is  a perfect fluid with
perturbed energy density and pressure,
$\rho(t,x^i)={\bar\rho}(t)+\d \rho(t,x^i)$,
$P(t,x^i)={\bar P}(t)+\d P(t,x^i)$,
 and four-velocity
 \beq
 \label{u}
 u^\mu={\bar
u}^\mu+\d u^\mu, \quad \d u^\mu=\{-A, v^i/a\}, \quad v_k=
\vec \nabla_k v+{ v}^V_k,
\eeq
where $v^V_i$ is transverse, $\vec \nabla_i v^V{}^i=0$.

\def\ab{\bar{\alpha}}
\def\rhob{\bar{\rho}}
\def\pb{\bar{P}}
\def\rhof{{\delta \rho}}
\def\pf{{\delta P }}
\def\Xf{{\delta X}}
\def\af{{\delta \alpha}}

It is easy to check that the covector
$\zeta_a$ automatically vanishes in the background. At
{\it linear order},  its spatial components are simply \cite{Langlois:2005ii,Langlois:2005qp}
\beq
\zeta_i=\partial_i\zeta, \qquad \zeta\equiv
\af-{H\over {\rhob}'}
\rhof,
\label{zeta1}
\eeq
where $\HH$ denotes the Hubble rate, $\HH =
a'/a={\ab}'$ and we recall that a prime denotes the partial derivative with
respect to the cosmic time $t$.
The quantity $\zeta$ defined above differs from the  familiar definition \cite{Bardeen:1983qw}
\beq
\zeta_B\equiv-\psi -\HH{\rhof\over\rhob'},
\eeq
because $\af$ differs from $-\psi$. They are related by \cite{Langlois:2005qp}
\begin{eqnarray}
\af&=& -\psi+{1\over 3}\nabla^2 E+ \frac{1}{3}\int
{d t}\, \frac{\nabla^2 v}{a}, \label{alpha}
\end{eqnarray}
and they coincide when  the gradient terms are negligible.

The  components of the non-adiabatic term $\Gamma_a=\partial_a
P-(\dot P/\dot\rho)\partial_a\rho$ can be deduced directly from
the components of $\zeta_a$ by substituting $P$ to $\alpha$.
One finds
\beq
\Gamma_i= \partial_i \Gamma, \qquad \Gamma \equiv
\pf-{{\pb}'\over {\rhob}'} \rhof.
\label{Gamma1}
\eeq

Putting all together 
Eq. (\ref{dot_zeta}) gives, at first order,
\beq
\zeta'=\zeta_B' +{1\over 3}\nabla^2(
E'+ v/a) =-{\HH\over \rhob+\pb} \left(\pf-{{\pb}'\over {\rhob}'} \rhof
\right).
\eeq
For adiabatic perturbations, the right-hand side vanishes and the  quantity $\zeta$ is
conserved at {\it all scales} because it comes directly from the covector  $\zeta_a$. By contrast, $\zeta_B$ is conserved only when the gradient terms are negligible.

\subsubsection{Second-order perturbations and beyond} \label{Second}

\def\as{{\delta \alpha}^{(2)}}
\def\Xs{{\delta X}^{(2)}}
\def\rhos{{\delta \rho}^{(2)}}
\def\ps{{\delta P}^{(2)}}

The covariant approach turns out very useful to go beyond linear order in perturbation theory. 
Any function $X$ can be decomposed in the form
\beq
X(t,x^i)={\bar X}(t)+ \Xf^{(1)}(t,x^i)+
\Xs(t,x^i) + \Xf^{(3)}(t,x^i) + \ldots \ ,
\eeq
where a quantity with a superscript ${}^{(n)}$ is the $n$-th order contribution to $X$ in the perturbation theory expansion \cite{Bruni:1996im} (note that here we do not follow the convention of including a numerical factor $1/n!$ in front of the $n$-th contribution). 

Expanding 
$\zeta_i=\partial_i\alpha-(\dot\alpha/\dot\rho)\partial_i\rho$ at second order one finds that, 
in contrast with the first order expression, 
$\zeta^{(2)}$ cannot be written as the gradient of a scalar perturbation.
Indeed, after some manipulations one finds, at second order \cite{Langlois:2005ii,Langlois:2005qp},
\beq
\label{zeta2} \zeta_i^{(2)}
=\partial_i\zeta^{(2)}
+ {\rhof \over
\rhob'}\partial_i \zeta^{(1)}{}',
\eeq
with
\beq
\zeta^{(2)} 
=\as - {H\over {\rhob}'}\rhos
 - \frac{1}{{\rhob}'} {\af}' \rhof+ \frac{H}{{{\rhob}'}{}^2} {\rhof}
 {\rhof}'+\frac{1}{2 {\rhob}'}
 {\left({H\over {\rhob}'}\right)}'
 {\rhof}^2.
  \label{zeta_scal}
\eeq
On large scales, using 
\beq
\alpha \approx \ln a - \psi - \psi^2, \label{relation_added}
\eeq
one can show that $\zeta^{(2)}$ is directly related  to the conserved second-order
quantity defined by Malik and Wands \cite{Malik:2003mv}:  $\zeta^{(2)} 
\approx \zeta_{\rm MW}^{(2)}
- (\zeta_{\rm MW}^{(1)})^{2}
$ (see also the discussion in \cite{Lyth:2005du}). We use the symbol $\approx$ to denote an equality only valid on large scales.

Now that we have identified a second-order expression
for $\zeta$, we can expand Eq.~(\ref{dot_zeta}) at second order.
On making use of the second line of Eq.~(\ref{Lie_def}) to reexpress the Lie
derivative along $u^a$ in terms of the coordinate time derivative, and
retaining only terms at second order, we have
\beq
\dot \zeta_i^{(2)} 
=  \zeta_i^{(2)}{}' 
-  A
\zeta^{(1)}_i{}' +\left( \frac{v^j}{a} \partial_j \zeta^{(1)}_i +
\zeta^{(1)}_j
\frac{\partial_i v^j}{a} \right) .
\eeq
Finally, on making use of Eqs.~(\ref{zeta1}), (\ref{zeta2}), and
(\ref{zeta_scal}), and that $\Theta = 3 (1-A) \alpha'$ up to
first order, we can explicitly write the conservation equation
(\ref{dot_zeta}) {\em up to  second order} and {\em on all scales}:
\beq
  \zeta^{(2)}{}' 
  = - \frac{\HH}{\bar \rho+ \bar P} \Gamma^{(2)} 
  -\frac{1}{\bar \rho+ \bar P} \Gamma^{(1)} 
  \zeta^{(1)}{}'  - 
\frac{v^j}{a}
\partial_j \zeta^{(1)} . \label{conserv_zeta2}
\eeq
The definition of $\Gamma^{(2)}
$ can be read from the second-order expression of
$\zeta^{(2)}
$ by substituting $P$ to $\alpha$. For adiabatic
perturbations, we find that at second order the scalar variable
$\zeta
$ is conserved only on large scales, when the last
term on the right-hand side of Eq.~(\ref{conserv_zeta2}) can be
neglected.

\def\z{\zeta}
\def\r{\rho}
\def\rb{\bar{\rho}}
\def\a{\alpha}
\newcommand{\nn}{\nonumber}

One can  extend straightforwardly the
procedure presented here to higher orders in the perturbation
expansion \cite{Enqvist:2006fs,Lehners:2009ja}. Expanding Eq.~(\ref{zeta_a}) at third order, after some
manipulations one can write \beq  \z_i^{(3)} 
= \partial_i  \z^{(3)} +\frac{\delta \r}{\rb'}\partial_i \z^{(2)'}
+\frac{\delta \r^{(2)}
}{\rb'}
\partial_i \z^{(1)'} 
-\frac{\rb''}{2\rb'^3}\delta
\r^2\partial_i \z^{(1)'} 
+ \frac{\delta
\r^2}{2\rb'^2}\partial_i \z^{(1)''},
\label{zetai3}
 \eeq
with
\beq 
\begin{split}
\z^{(3)} = & \delta \a^{(3)}-\frac{H}{\rb'}\delta
\r^{(3)}-\frac{1}{\rb'}\left(\delta \a'- \frac{H}{\rb'}\delta
\r'\right)\left(\delta \r^{(2)}-\frac{\delta \r \,\delta
\r'}{\rb'}+\frac{\rb'' \delta \r^2}{2 \rb^{'2}} \right)
-\frac{\delta \r}{\rb'} \left( \delta \a^{(2)'}-
\frac{H}{\rb'}\delta \r^{(2)'} \right)\\ + &   \frac{\delta
\r^2}{2 \rb^{'2}} \left(\delta \a''- \frac{H}{\rb'}\delta \r''
\right)+\frac{1}{\rb'}\left(\frac{H}{\rb'} \right)' \delta
\r^{(2)} \delta \r-\frac{1}{\rb^{'2}}\left(\frac{H}{\rb'}
\right)' \delta \r^2 \delta \r'-\frac{1}{6 \rb'} \left[
\frac{1}{\rb'} \left(\frac{H}{\rb'} \right)'\right]'\delta
\r^3. \label{zeta3-explicit} 
\end{split}
\eeq
Its evolution equation {\em up to third order}
can be found by expanding Eq.~(\ref{dot_zeta}). 
On large scales it reads
 \cite{Lehners:2009ja}
\beq
\z^{(3)}{}'
\approx
-\frac{H}{\bar \rho+\bar P}\Gamma^{(3)}
-\frac{1}{\bar \rho+\bar P} (\Gamma^{(1)}
\z^{(2)}{}' + \Gamma^{(2)}
\z^{(1)}{}' )
-\frac{1}{3H(\bar\rho+\bar P)}\Gamma
(\z')^2,
\label{zeta3'} \eeq 
where $\Gamma^{(3)}
$ is defined in the same
way as $\zeta^{(3)}
$ in Eqs.~(\ref{zetai3}) and
(\ref{zeta3-explicit}) with the replacements $\a \to P, \,
\zeta \to \Gamma$.

In \cite{Enqvist:2006fs} one can find a discussion on perturbations beyond the third order, as well as  a general expression for a quantity $\zeta^{(n)}$, defined at any order $n$, which is conserved on large scales  for adiabatic perturbations.

\subsubsection{The issue of gauge-invariance}
One can check that, on large
scales, $\zeta^{(2)}$ behaves as a gauge-invariant
quantity. Given a vector $\xi^a = \sum_n \frac{1}{n!} \,
\xi_{(n)}^a$, the gauge
transformation it generates is defined by the transformation
law of tensors  \cite{Malik:2008im} (whose coordinate functions
are a particular case),
\beq
 {\bf \tilde T} \rightarrow e^{{\cal L}_{\xi}} {\bf T}. \label{gauge_transf}
\eeq
Since $\zeta_a$ vanishes at zeroth order, $\zeta_a$ is
automatically gauge-invariant at {\em first order}, according to
the first expression above. However, $\zeta_i$ is not
gauge-invariant at second order and
the corresponding gauge transformation is given by
\beq
\zeta_i^{(2)}\rightarrow \zeta_i^{(2)}+ {\cal L}_{\xi_{(1)}}
\zeta_i^{(1)} \approx \zeta_i^{(2)}+ \xi_{(1)}^0\partial_0\zeta_i^{(1)},
\eeq
where for the last equality we have neglected the terms of higher
order in spatial gradients, concentrating on large scales. 
By noting that at first order $ {\rhof}/{\rhob'} \rightarrow {\rhof}/{\rhob'}+ \xi^0_{(1)}$, 
it is easy to see using Eq.~(\ref{zeta2}) that $\partial_i \zeta^{(2)}$, or equivalently $\zeta^{(2)}$,
is gauge-invariant at second order, on large scales. 
This can also be checked by directly applying the gauge transformation (\ref{gauge_transf}) on $\z^{(2)}$ defined in Eq.~(\ref{zeta_scal}). This analysis can be repeated at third \cite{Lehners:2009ja} or any higher order $n$ \cite{Enqvist:2006fs} and one can show that the quantity $\z^{(n)}$ is gauge invariant on large scales.

\subsection{Dissipative and interacting fluids} 
It is possible to extend our covariant approach to the case of dissipative and interacting fluids, as discussed in detail in \cite{Langlois:2006iq}.
In that case, the energy-momentum tensor is more complicated,
\beq
T_{ab}=\rho u_a u_b+ P h_{ab}+ q_{a} u_{b}+q_b
u_a +\pi_{ab}, \label{emt2}
\eeq
where the  energy flow $q_a$ and anisotropic stress $\pi_{ab}$ satisfy 
\beq
q_au^a=0, \quad \pi_{ab}=\pi_{ba}, \quad \pi_a^{\ a}=0, \quad \pi_{ab}u^b=0.
\eeq
The continuity equation in this case reads
 \beq
 \label{continuity2}
\dot\rho + \Theta (\rho + P)= {\cal D}, \quad \quad {\cal
D}\equiv-\left(\pi^{ab} \sigma_{ab} + D_a q^a + 2 q^a a_a\right),
\eeq
where $D_a$ denotes  the spatially projected covariant derivative, orthogonal to the four-velocity
$u^a$. For a generic tensor  ${\cal T}$, the definition is (see e.g. \cite{Ellis:1989jt})
\beq
D_a {\cal T}_{b\dots}^{\ c\dots}\equiv h_a^{\ d}h_b^{\ e}\dots h^{\
c}_f\dots\nabla_d {\cal T}_{e\dots}^{\ f\dots}.
\eeq
As a consequence of the presence of the dissipative terms on the right-hand side of (\ref{continuity2}), the evolution equation for $\zeta_a$ is slightly more involved and now reads
\beq
\dot\zeta_a= \frac{3 \dot \alpha^2}{ \dot \rho}
\left(\Gamma_{a}+\Sigma_{a}\right) , \label{conserv2}
\eeq
where 
\beq
 \label{Sigma}
 \Sigma_a\equiv
D_a \beta- {\dot \beta\over \dot\rho}D_a\rho\, , \qquad  \beta \equiv -\frac{\cal D}{\Theta}\, .
\eeq
This means that dissipative terms affect the evolution of the covector $\zeta_a$ in addition to the non-adiabatic term.

\section{Cosmological scalar fields}

\def\Tdot#1{{{#1}^{\hbox{.}}}}
\def\Tddot#1{{{#1}^{\hbox{..}}}}
\def\D{{\cal D}}
\def\d{{\delta}}
        \def\T{{\bf T}}
\def\u{{\partial_u}}
\def\uc{u}
\def\qc{q}
\def\Dc{D}
\def\sigmad{\dot \sigma}
\def\phid{\dot \phi}
\def\chid{\dot \chi}
\def\varphid{\dot \varphi}
\def\L{{\cal L}}
\def\R{{\cal R}}
\def\HH{{\cal H}}
\def\hc{h}
\def\I{I}
\def\It{{(I)}}
\def\J{J}
\def\Jt{{(J)}}
\def\p{\varphi}
\def\vp{{\boldsymbol{\varphi}}}
\def\e{{e}}

\def\Xf{X^{(1)}}
\def\Xs{X^{(2)}}
\def\af{{\delta \alpha}}
\def\as{{\delta \alpha}^{(2)}}
\def\sif{{\delta \sigma}}
\def\sis{{\delta \sigma}^{(2)}}
\def\sf{{\delta s}}
\def\ss{{\delta s}^{(2)}}
\def\phf{{\delta \varphi}}
\def\phs{{\delta \varphi}^{(2)}}
\def\phif{{\delta \phi}}
\def\chif{{\delta \chi}}
\def\sif{{\delta \sigma}}
\def\phis{{\delta \phi}^{(2)}}
\def\rhof{{\delta \rho}}
\def\rhos{{\delta \rho}^{(2)}}
\def\ps{{\delta P}^{(2)}}
\def\pf{{\delta P}}
\def\ab{\bar{\alpha}}
\def\rhob{\bar{\rho}}
\def\pb{\bar{P}}
\def\sib{\bar{\sigma}}
\def\phib{\bar{\phi}}
\def\chib{\bar{\chi}}
\def\sb{\bar{s}}
\def\Xb{\bar{X}}
\def\Xdb{\bar{\dot X}}
\def\thetab{\bar{\theta}}
\def\tackr{&\!\!\!}
\def\tackl{&\!\!\!}

We now consider the situation where matter is composed of several scalar fields, concentrating on the case \cite{Langlois:2006vv}. We study $N$ scalar fields
minimally coupled to gravity with Lagrangian density
\beq
{\cal L} = - \frac{1}{2}
\partial_a \p_\I \partial^a \p^\I -  V(\p_K) ,
\label{lagrangian}
\eeq
where $V$ is the potential and the summation
over the field indices ($I,J,\ldots$)  is implicit. 
For simplicity we assume here
canonical kinetic terms but the nonlinear extension to a large class of models with non-canonical kinetic terms, such as those studied at the linear level in \cite{Langlois:2008mn}, has also been developed in \cite{RenauxPetel:2008gi}. 
The
energy-momentum tensor derived from the above Lagrangian reads
\beq
T_{ab}= \partial_a \p_\I \partial_b \p^I -\frac{1}{2} g_{ab}
\left( \partial_c \p_\I
\partial^c \p^I + 2 V \right). \label{EMT}
\eeq
Given an {\it arbitrary} unit timelike vector field $u^a$, it is always
possible to decompose this energy momentum tensor in the form (\ref{emt2})
with, in our case, 
\bea
\rho \tackl = \tackr \frac{1}{2} \left(   \dot \p_I\dot\p^I + \Dc_a
\p^I \Dc^a \p_\I \right) +V , \qquad 
P =  \frac{1}{2}  \left(
\dot \p_I\dot\p^I - \frac{1}{3} \Dc_a \p_\I
\Dc^a \p^I \right) -V, \label{Ptot}\\
q_a \tackl = \tackr  - \dot \p_\I \Dc_a \p^\I, \qquad
\pi_{ab} =    \Dc_a
\p_\I \Dc_b \p^I -\frac{1}{3} \hc_{ab} \Dc_c \p_\I \Dc^c \p^\I .
\label{as}
\eea

The evolution equations for the scalar fields are the  $N$ Klein-Gordon
equations given by
\beq
-\nabla_a\nabla^a\p_\I+\frac{\partial V}{\partial \p_\I}=
\ddot \p_\I + \Theta \dot \p_\I +V_{,\varphi_\I}-\Dc_a \Dc^a \p_\I - a^a
\Dc_a \p_\I =0, \label{evolp}
\eeq
where the second expression is obtained by using  a decomposition
into (covariant) time-like and space-like gradients defined with respect
to $u^a$.

\subsection{Adiabatic and entropy covectors}

For simplicity we now
restrict our analysis to the case of two scalar fields, which we
will denote by $\phi$ and $\chi$. In the two-field case it is possible to introduce a particular
basis in the field space in which various field dependent
quantities are decomposed into so-called adiabatic and entropy
components. In the linear theory this decomposition
 was first introduced in
\cite{Gordon:2000hv} for two fields. This decomposition is particularly convenient  to follow the time evolution of the curvature perturbation, which is sourced by the entropic perturbations.  For the multi-field case it
is discussed in \cite{GrootNibbelink:2001qt} in the linear theory and in
\cite{Rigopoulos:2005xx} in the nonlinear context.

In our case, the corresponding basis consists, in the
two-dimensional field space, of   a unit vector $\e_\sigma^I$
defined in the direction of the velocity of the two fields, and thus
{\em tangent} to the trajectory in field space, and of a  unit
vector $\e_s^I$ defined along  the direction {\em orthogonal} to
it, namely
\beq
\e_\sigma^\I \equiv  
({\dot \phi}/{\dot\sigma}, {\dot \chi}/{\dot\sigma} )=\left(\cos \theta,\sin \theta  \right),
\qquad \e_s^\I \equiv
(- {\dot \chi}/{\dot\sigma}, {\dot \phi}/{\dot\sigma})=\left(-\sin \theta , \cos \theta \right),
 \label{e1}
\eeq
where we have introduced 
the {\em formal} notation
$\sigmad\equiv(\dot\phi^2+\dot\chi^2)^{1/2}$.
Note that this notation can be misleading as, in
general, in the nonlinear context $\sigmad$ {\em is not} the derivative along $u^a$ of a
scalar field $\sigma$.

The  angle $\theta$, which characterizes the rotation angle between the initial field basis and the adiabatic/entropic basis,   is here an
inhomogeneous  quantity which depends on  time and space. By
taking the time derivative of the basis vectors $\e_\sigma^\I$ and
$\e_s^\I$, we get
\beq
\dot \e_\sigma^\I =  \dot \theta \e_s^\I  , \qquad
\dot \e_s^\I = - \dot \theta
\e_\sigma^\I. \label{angle_1}
\eeq

Making use of the basis (\ref{e1}), one can then introduce  two
linear combinations of the scalar field gradients and thus define
two covectors, respectively denoted by $\sigma_a$ and $s_a$, as \cite{Langlois:2006vv}
\bea
\sigma_a \tackl \equiv \tackr \e_\sigma^\I \partial_a \p_\I=\cos\theta\, \partial_a\phi+\sin\theta\, \partial_a\chi
\label{tan_ort1}, \\
 s_a \tackl \equiv \tackr \e_s^\I \partial_a \p_\I  =-\sin\theta\, \partial_a\phi+\cos\theta\, \partial_a\chi
 \label{tan_ort2}.
\eea
We will call these two covectors the {\em adiabatic} and {\em
entropy}  covectors, respectively, by analogy with the similar
definitions in the linear context \cite{Gordon:2000hv}. Whereas
the entropy covector $s_a$ is orthogonal to the four-velocity
$u^a$, i.e., $u^a s_a = 0$, this is not the case for $\sigma_a$
which contains a ``longitudinal'' component:
$u^a\sigma_a=\dot\sigma$. It is also useful to introduce the
spatially projected version of (\ref{tan_ort1}) and (\ref{tan_ort2}),
\def\sp{\sigma^{_\perp}}
\beq
\sp_a \equiv \e_\sigma^\I D_a \p_\I = \sigma_a+
\dot\sigma u_a \, , \qquad
s^\perp_a \equiv
\e_s^\I D_a \p_\I = s_a\, .
\label{perp}
\eeq

The ``adiabatic'' combination of the
Klein-Gordon equations, i.e., the contraction of (\ref{evolp}) by
$\e_\sigma^\I$, yields 
\beq
\ddot \sigma + \Theta \dot \sigma + V_{,\sigma}= \nabla^a \sp_a
  -Y_{(s)},
\label{evolsigma}
\eeq
where we have defined
\beq
V_{, \sigma} \equiv \e_\sigma^\I V_{,\p_\I}, \qquad Y_{(s)}\equiv \frac{1}{\dot \sigma} (\dot s_a + \dot \theta \sigma_a^\perp )s^a\, .
\label{Y_s}
\eeq
Since $\e_s^\I \ddot \p_\I=\dot\theta\dot\sigma$, the ``entropic'' combination of the
Klein-Gordon equations, i.e., the contraction of (\ref{evolp})
with $\e_s^\I$, gives simply
\beq
\dot \sigma \dot \theta + V_{,s}= \nabla_a s^a
  + Y_{(\sigma)}, \label{thetadot}
\eeq
with 
\beq
V_{, s} \equiv \e_s^\I V_{,\p_\I}, \qquad 
Y_{(\sigma)}\equiv \frac{1}{\dot \sigma} (\dot s_a + \dot \theta
\sigma^\perp_a )\sp{}^a\, .
\label{Y_sigma}
\eeq

In summary,  the Klein-Gordon
equations for the fields $\phi$ and $\chi$ have been replaced by  the equations (\ref{evolsigma}) and
(\ref{thetadot}), whose left-hand side  has
exactly the same form as the homogeneous equations in a Friedmann
universe. However, these equations  capture the fully nonlinear dynamics of the scalar fields and
their right-hand side contains nonlinear (quadratic) terms represented by $Y_{(s)}$ and
$Y_{(\sigma)}$, sourcing the adiabatic and entropy equations
respectively.

We now consider the evolution equations for the covectors $\sigma_a$ and
$s_a$, or rather for their spatial projections.
The  adiabatic evolution equation is given by
\bea
&&(\ddot \sigma_a)^\perp + \Theta (\dot \sigma_a)^\perp
+\dot\sigma D_a\Theta +\left(V_{,\sigma
\sigma}+\dot \theta \frac{V_{,s}}{\dot \sigma} \right)
\sigma_a^\perp -D_a \left(\nabla^c \sp_c\right)
=\left(\dot \theta-\frac{V_{,s}}{\dot \sigma}\right) \dot s_a
  \nonumber \\
&& \qquad  +\left(\ddot\theta - V_{,\sigma s}+ \Theta
\dot \theta\right) s_a -
D_aY_{(s)}\ , \label{sigma1_spatial}
\eea
where we have introduced the notation
\beq
V_{,\sigma \sigma} \equiv \e_\sigma^\I \e_\sigma^\J V_{,\p_\I
\p_\J }, \qquad V_{,s s} \equiv \e_s^\I \e_s^\J V_{,\p_\I \p_\J },
\qquad V_{,s \sigma} \equiv \e_s^\I \e_\sigma^\J V_{,\p_\I \p_\J}\, ,
\eeq
for the second derivatives of the
potential. Note that the spatial projection and the time derivative do not commute for $\sigma_a$.
The entropic evolution equation  reads
\bea
\ddot s_a - \frac{1}{\dot \sigma}(\ddot \sigma + V_{,\sigma})
\dot s_a +( V_{,ss}-\dot \theta^2 ) s_a
-D_a \left(\nabla_c s^c\right)
=- 2
\dot \theta (\dot \sigma_a)^\perp   
+\left[\frac{\dot
\theta}{\dot \sigma} (\ddot \sigma+V_{,\sigma})
-\ddot \theta - V_{,\sigma s}\right] \sigma_a^\perp+
D_a Y_{(\sigma)}\ ,\label{s1}
\eea
where  the covectors $\dot s_a$ and
$\ddot s_a$ are purely spatial, i.e. $(\dot s_a)^\perp =
\dot s_a $ and $(\ddot s_a)^\perp = \ddot s_a $.

Starting from the fully nonlinear Klein-Gordon equations, we have
thus managed to obtain a system of two coupled equations
(\ref{sigma1_spatial}) and (\ref{s1}), which govern the evolution
of our nonlinear adiabatic and entropy components \cite{Langlois:2006vv}.  Remarkably, they
are rather simple and they look very similar to the linear equations  for the perturbations
$\delta\sigma$ and $\delta s$ (see \cite{Gordon:2000hv}). Furthermore, since these equations are exact, it is straightforward to  expand them up to second or higher orders, as we  will
show explicitly in Sec.~\ref{sec:coord_fields}.

\subsection{Generalized curvature perturbations and large-scale evolution}
\label{sec:ls}
Let us now turn to the generalized curvature perturbation $\zeta_a$, which we introduced in Eq~(\ref{zeta_a}) for a cosmological fluid. 
Although a system of several scalar fields cannot be described as a perfect fluid, its energy-momentum tensor is of the form (\ref{emt2}) with  
\bea
\rho \tackl = \tackr \frac{1}{2} \left(   \dot \sigma^2 + \Pi  \right) +V ,\label{EMTPrho} \qquad
P  =  \frac{1}{2} \left(   \dot \sigma^2 -\frac13 \Pi  \right) -V ,
\label{EMTP} \\
q_a \tackl = \tackr -\dot\sigma \sp_a, \label{qa} \qquad
\pi_{ab}  =     \Pi_{ab} -\frac13 h_{ab} \Pi, \label{pi}
\eea
where we have defined
\beq
\Pi_{ab} \equiv \sp_a \sp_b +s_a s_b,  \qquad \Pi \equiv \sp_c
{\sp}^c +s_c s^c\, .
\eeq
The quantity $\zeta_a$ for the two-fluid system therefore satisfies an evolution equation of the form (\ref{conserv2}).

In the linear theory, an alternative quantity to describe the primordial perturbation is the comoving curvature perturbation. A natural  nonlinear extension of this quantity in our formalism is (see also \cite{Pitrou:2007xy} for an alternative definition)
\beq
\R_a \equiv   - D_a \alpha  -  \frac{\dot \alpha}{(\dot \varphi_J
\dot \varphi^J)} q_a \;. \label{R_N} 
\eeq
As in the linear case, this expression can be directly related to the analog of the Mukhanov-Sasaki variables \cite{Sasaki:1986hm,Mukhanov:1988jd}, if we generalize the latter by defining 
\beq
Q^I_a \equiv D_a \varphi^I - \frac{\dot \varphi^I}{\dot \alpha}
D_a \alpha 
\eeq
for each scalar field. 
The comoving curvature perturbation $\R_a$ is then proportional to the {\em adiabatic}  combination of the $Q^I_a$, i.e. 
\beq
Q_a \equiv\e_{\sigma \I} Q^I_a = \sigma_a^\perp - \frac{\dot \sigma}{\dot \alpha} D_a
\alpha  , \label{Q_two}
\eeq
since one can write
$\R_a = {\dot \alpha}{}\  \dot
\varphi_I Q^I_a / (\dot \varphi_J \dot \varphi^J)$.

Comparing the definitions of  $\zeta_a$ and $\R_a$ shows that they satisfy
the relation 
\beq
\zeta_a + \R_a = - \frac{\dot \alpha}{\dot \rho} \epsilon_a, \qquad
\epsilon_a\equiv\Dc_a\rho- \frac{\dot \rho}{\dot \sigma}\sp_a,
\label{epsilon}
\eeq
where the covector $\epsilon_a$ can be interpreted as a covariant generalization of the {\em comoving
energy density} perturbation.

Let us now concentrate on the {\it large scale limit}  by neglecting terms of higher order in  spatial gradients.
Using the  {\em energy constraint}  $u^a G_{ab}u^b=8\pi G \rho$ and the 
{\em momentum constraint}
$u^b G_{bc}h^c_a= 8\pi G q_a$, it can be shown that  the comoving energy density
perturbation, $\epsilon_a$, defined  in Eq.~(\ref{epsilon}), can be
neglected on large scales if the shear can also be neglected in
this limit. This is the case  in an expanding perturbed FLRW universe, where the shear rapidly decreases on large scales. 

Then, from their definition (\ref{perp}),
$\sigma_a^\perp$ and $s_a$ are first-order quantities with
respect to spatial gradients while the scalars $Y_{(s)}$ and $Y_{(\sigma)}$ are
second order, since they are
quadratic in $\sigma_a^\perp$ and $s_a$ (or their time
derivatives). Hence, the right-hand side of Eq.~(\ref{evolsigma})
and of Eq.~(\ref{thetadot}) can be neglected on large scales.
Taking into account these simplifications and neglecting on large scales $\epsilon_a$ and the terms containing $\Pi$, one finds that the evolution equations in Eqs.~(\ref{sigma1_spatial}) and (\ref{s1}) simplify to give 
\beq
(\ddot \sigma_a)^\perp + \Theta (\dot \sigma_a)^\perp +\dot\sigma
D_a\Theta +(V_{,\sigma \sigma}-\dot \theta^2)
\sigma_a^\perp  \approx 2 \Tdot{(\dot \theta s_a )} -2
\dot \theta \frac{V_{,\sigma}}{\dot \sigma}
 s_a
, \label{si_evol_ls}
\eeq
and
 \beq 
\ddot s_a +\Theta \dot s_a + ( V_{,ss}+3 \dot \theta^2
) s_a \approx 0. \label{s_evol_ls}
\eeq
While the entropy mode appears as a source term in the adiabatic equation, the entropy  equation involves only the covector $s_a$.

On large scales, neglecting $\epsilon_a$ in  Eq.~(\ref{epsilon}) implies that  the
 uniform density  and comoving perturbations $\zeta_a$ and $\R_a$ {\em coincide} (up
to a sign),
\beq
\zeta_a + \R_a \approx 0. \label{zeta_R_ls}
\eeq
Moreover, in the same limit, the source term in the evolution equation for $\zeta_a$, Eq.~(\ref{conserv2}), is dominated by the 
gradient of the potential $V$ along the entropic direction, so that 
 \beq \dot \zeta_a \approx - \frac{2}{3}  \frac{\Theta^2}{ \dot
\rho} V_{,s} s_a \label{zeta_evol_ls}.
\eeq
This equation is sufficient to describe the evolution of the adiabatic mode in the large scale limit,  making Eq.~(\ref{si_evol_ls}) redundant.
The analogs of the spatial components of Eqs.~(\ref{zeta_evol_ls}) and (\ref{s_evol_ls}) have also been derived in \cite{Rigopoulos:2005xx}.

\subsection{Link with the coordinate approach}
\label{sec:coord_fields}

\subsubsection{Linear theory}
\label{sec:firstorder_fields}

We now relate  our covariant approach with the more familiar
coordinate based formalism, starting with the linear theory. We use the metric and notation of Sec.~\ref{Linear} and
the scalar fields are decomposed into a background and a perturbed
part, $\p_\I(t,x^i)={\bar\p_\I}(t)+\d \p_\I(t,x^i)$.
We  now need to specify the components of the unit vector $u^a$, which defines the time
direction in our covariant approach. 
In contrast with the perfect fluid case, where $u^a$ has a physical meaning, the unit vector $u^a$ is here arbitrary. In a given coordinate system it is convenient  to   choose $u^\mu$ such that $u_i=0$ at first order.
This implies that the components
of $u^\mu$ and of the ``acceleration'' vector  are given respectively by
\beq
\label{components_u}
u^\mu=\{1-A, -B^i/a\} ,\qquad 
a^\mu=\{0, \vec\nabla^i A /a^2\}\ .
\eeq

The
background equations of motion can be deduced immediately from
Eqs. (\ref{evolsigma}) and (\ref{thetadot}) and read
\beq
{\sib}'' + 3H  {\sib}' +\bar V_{,\sigma}= 0, 
\qquad
\sib'  \thetab' + \bar V_{,s}= 0.
\eeq
From their definition, one finds that
the {\it spatial} components of $\sigma_a$   and $s_a$, at linear order, can be
expressed as
\bea
\label{dsigma}
\delta\sigma_i&=&\partial_i\delta\sigma, \qquad 
\sif\equiv
\cos\thetab \, \phif+\sin\thetab\, \chif,\\
\delta s_i&=&\partial_i \sf, \qquad
\sf\equiv \cos\thetab \,    \chif-\sin\thetab\, \phif,
\eea
which  coincide with the notations of \cite{Gordon:2000hv}. 

Linearizing the evolution equations for $\sigma_a$ and
$s_a$, eqs.~(\ref{sigma1_spatial}) and (\ref{s1}), one easily obtains the linearized equations for
$\delta\sigma$ and $\delta s$. The latter can be written as 
\beq
\sf'' +3H\sf' +( \bar V_{,ss}+3 \thetab'{}^2 )
\sf-\frac{1}{a^2}\vec\nabla^2\sf =- 2 \frac{\thetab'}{\bar
\sigma'} \delta \epsilon\, ,
\label{s_evol_1}
\eeq
where on the right-hand side appears  the first-order
comoving energy density perturbation $\delta\epsilon$, defined by
\beq
\delta \epsilon_i = \partial_i \delta \epsilon, \qquad \delta
\epsilon \equiv \delta \rho - \frac{\rhob'}{\sib'} \sif ,
\eeq
which follows from  the definition (\ref{epsilon}) of $\epsilon_a$.
The quantity $\delta\epsilon$ can in general be neglected on large scales, since it is second order in
 spatial gradients,  as can be seen from  the
relativistic Poisson-like equation.

Let us now turn to the adiabatic equation. Since 
 $\sif$ is not gauge-invariant, in contrast with $\sf$,  it is more useful to consider the gauge invariant
Sasaki-Mukhanov variable $Q_{\rm SM}$, defined
 as \cite{Sasaki:1986hm,Mukhanov:1988jd}
\beq
Q_{\rm SM} \equiv \sif + \frac{\sib' }{H} \psi.
\eeq
Note that the above traditional definition coincides with the scalar  quantity $Q$ that can be extracted from our definition
  of  $Q_a$ given earlier in
Eq.~(\ref{Q_two}), via $Q_i = \partial_i Q$, only in the large scale limit, because $\psi$ and $-\af$ coincide only in this limit.
The evolution equation of $Q_{\rm SM}$ reads
\cite{Taruya:1997iv,Gordon:2000hv}
\beq
Q_{\rm SM}''+ 3H Q_{\rm SM}' + \left[\bar V_{, \sigma \sigma} -
\thetab'{}^2 - 2\frac{H'}{H}  \left(
  \frac{\bar V_{, \sigma}}{\sib'} +
\frac{H'}{H} - \frac{\sib''}{\sib'}\right) - \frac{\vec \nabla^2}{a^2} \right] Q_{\rm SM}  = 2 (\thetab' \sf)' -2 \thetab'\left(\frac{\bar V_{,
\sigma}}{\sib'} + \frac{H'}{H}  \right) \sf. \label{equation_Q_1}
\eeq 
On  {\em large scales} the adiabatic evolution is simpler and governed by 
the first integral
\beq
Q_{\rm
SM}' + \left( \frac{H'}{H} - \frac{\sigma''}{\sigma'} \right) Q_{\rm SM}
- 2 \thetab' \sf \approx 0\, ,
\label{integral_Q_1}
\eeq 
which simply expresses that $\delta\epsilon$, written in terms of $Q_{\rm SM}$ and $\delta s$, is negligible on large scales. 
One can easily check that the large-scale limit of
(\ref{equation_Q_1}) follows from  the first
integral (\ref{integral_Q_1}).

Let us now consider  the evolution equation for $\zeta_a$. The  spatial
components of $\zeta_a$, at linear order, are given by eq.~(\ref{zeta1}) as for the fluid case: $\zeta_i\equiv \partial_i\zeta$. Similarly, one can define $\R$ by $\R_i=\partial_i\R$. Note that $\R$ coincides with the familiar comoving curvature perturbation only in the large scale limit, because of the difference between $\alpha$ and $\psi$, as already discussed for $\zeta$. 
Eq.~(\ref{zeta_R_ls}) implies 
a simple relation between $\zeta$ and $\R$,
\beq
\zeta+\R=-\frac{\ab'}{\rhob'} \delta \epsilon\, ,
\label{relation_R_zeta}
\eeq
which shows that $\zeta$ and $-\R$ coincide on large scales.

The evolution equation for $\zeta$ on large scales  follows from eq.~(\ref{zeta_evol_ls})
which, after  linearization, yields 
\beq
\zeta' \approx -
\frac{2H}{\sib'}\bar \theta' \delta s.
\eeq
One thus recovers   the familiar linear result~\cite{Gordon:2000hv} that the entropy perturbation is sourcing the evolution of the uniform density curvature perturbation on large scales.

\subsubsection{Second-order perturbations}
\label{sec:secondorder_fields}

As for the perfect fluid case in section \ref{Second},  we now expand the equations governing
$\sigma_a$, $s_a$ and $\zeta_a$ at second order in the
perturbations. 
For
simplicity, we will restrict ourselves to {\em large scales} and
we will thus start from the equations expanded in spatial
gradients discussed in Sec.~\ref{sec:ls}.

The second-order evolution equations for the perturbations of a {\em single}
scalar field, in a coordinate based approach, has been considered
in several references (see for instance
\cite{Acquaviva:2002ud,Noh:2004bc,Bartolo:2004if,Vernizzi:2004nc}).
The multi-field case has been first studied
by Malik in \cite{Malik:2005cy,Malik:2006ir} and, using the separate
universe approach, in \cite{Lyth:2005fi,Vernizzi:2006ve}. Here, we obtain directly
 the second-order evolution equations in terms of  the adiabatic and entropic
perturbations.

By expanding Eq.~(\ref{tan_ort1}) and  Eq.~(\ref{tan_ort2}) up to  second order, one can write~\cite{Langlois:2006vv}
\bea
\label{sigi2} \delta \sigma_i^{(2)} \tackl = \tackr \partial_i
\sis + \frac{\thetab'}{\sib'} \sif\partial_i \sf -\frac{1}{\sib'} V_i, \label{si_i} \\
\label{s_i}  \delta s_i^{(2)}  \tackl = \tackr \partial_i \ss +
\frac{\sif}{\sib'}  \partial_i \sf',
\eea
with
\bea
\label{si2} \sis \tackl \equiv \tackr \frac{\phib'}{\sib'} \phis +
\frac{\chib'}{\sib'} \chis + \frac{1}{2 \sib'} \sf \sf',  \\
\ss \tackl \equiv \tackr - \frac{\chib'}{\sib'} \phis +
\frac{\phib'}{\sib'} \chis -\frac{\sif}{\sib'}  \left( \sf' +
\frac{\bar \theta'}{2}
 \sif\right),\label{sis}
\eea
and where we have defined the spatial vector
\beq
V_i \equiv \frac{1}{2} (\sf \partial_i \sf' - \sf' \partial_i \sf) \, . \label{V_def}
\eeq
The definition of $\ss$ is chosen such that it is gauge invariant on large scales. 
 Since the adiabatic component $\sigma_a$ does not vanish at zeroth order, $\sis$ is
not a gauge invariant variable and can be chosen for convenience. Its definition here is 
such that the momentum perturbation almost vanishes (up to $V_i$) when $\sis=0$.

The presence of the spatial vector $V_i$ on the right-hand side of Eq.~(\ref{si_i}) is due to the fact that $\sigma_a$ defined in Eq.~(\ref{tan_ort1}) is not hypersurface orthogonal. Indeed, one can check using its definition that $\sigma_{[a} \nabla_b \sigma_{c]} = \dot \sigma^{-1} \sigma_{[a} \dot s_b s_{c]}$. 
As the momentum $\delta q_i$ is proportional to $\delta \sigma_i^\perp$, Eq.~(\ref{qa}), this implies that if $V_i$
does not vanish one cannot define at second-order a comoving
gauge,  i.e., such that $\delta q_i^{(1)}=0$ {\em and} $\delta
q_i^{(2)}=0$, in  contrast with the linear theory or the
single-field case \cite{Rigopoulos:2005xx}. 
However, the  evolution equation for $V_i$ on
large scales reads
\beq
V_i'+ 3 H V_i \approx 0,
\label{evolution_Vi}
\eeq
which implies that in an expanding universe $V_i$ decays
as $a^{-3}$ and rapidly become negligible even if it is nonzero
initially. Consequently, in the following we will ignore $V_i$ {\em on large
scales}.

Since $\sis$ is not gauge-invariant, it is useful to consider the Sasaki-Mukhanov
variable at second order $Q^{(2)}_{\rm SM}$, which on large scales coincides with the scalar quantity $Q^{(2)}$ which can be extracted by expanding at second order $Q_a$ defined in Eq.~(\ref{Q_two}),
\beq
Q_i^{(2)} = \partial_i \Qs + \frac{\af}{H}  \partial_i \Qf'
+  \frac{\thetab'}{\sib'} \Qf\partial_i \sf -\frac{1}{\sib'}V_i. \label{Q2_i}
\eeq
The Sasaki-Mukhanov $Q^{(2)}_{\rm SM}$ can be found by replacing Eq.~(\ref{relation_added}) 
in the definition of $Q^{(2)}$, which yields
\beq
\Qs_{\rm SM} \equiv \sis + {\sib'\over H}(\psi^{(2)} + \psi^2)
 + \frac{\psi}{H} \left[ Q_{\rm SM}^{(1)}{}' - \frac{1}{2}
 {\left({\sib'\over H}\right)}' \psi - \thetab' \sf \right]
\label{Q_2}.
\eeq
Restricted to a single scalar field, this definition coincides with the one given in \cite{Malik:2005cy}.
By using the energy and momentum constraint equations at second order it is possible to derive a first integral for $\Qs_{\rm SM}$ \cite{Langlois:2006vv}, extending at second order Eq.~(\ref{integral_Q_1}). 

Here we will simply derive the large-scale evolution equation for $\zeta$ {\em at second order}, which is enough to describe the evolution of the adiabatic mode.
Expanding up to second order Eq.~(\ref{zeta_evol_ls}) yields
\beq
\zeta^{(2)}{}'\approx -\frac{H}{\sib'^2} \left[ 2\bar \theta'
\sib' \ss
 -  \left( \bar V_{,ss} + 4 \bar \theta'{}^2
\right)  \sf^2 
+ \frac{\bar V_{,\sigma}}{\bar \sigma'} \delta s \delta s'
 \right]. \label{zeta2_evol}
\eeq

It is also useful to express our results in terms of $\R_a$. The
spatial components of $\R_a$ can be decomposed as
\beq
\R_i^{(2)} = \partial_i \R^{(2)} + \frac{ \sif}{\sib'}
\partial_i \R^{(1)}{}' -  \frac{H}{\sib'{}^2}V_i , \label{R_2_i}
\eeq
with
\beq
\R^{(2)} \equiv-  \as + {H\over \sib'}\sis
 + \frac{\sif}{\sib'} \left[ -\R^{(1)}{}' + \frac{1}{2}
 {\left({H\over \sib'}\right)}'
 \sif  +  \bar \theta' \frac{H}{\sib'}
 \sf \right]
\label{R_2}.
\eeq
The last term in Eq.~(\ref{R_2_i}) comes from the fact that, in contrast to $\zeta_a$, $\R_a$ is defined in
terms of the spatial momentum which cannot be expressed in general
as a pure gradient. However, in an expanding universe this term can be neglected and  $\R^{(2)}$ coincides with the second-order comoving curvature perturbation defined in
\cite{Maldacena:2002vr,Vernizzi:2004nc}.

It is easy to derive a first-order (in time) evolution equation
for $\R^{(2)}$ by noting that $\zeta^{(2)}$ and $\R^{(2)}$ are related on large scales. Indeed, by using Eqs.~(\ref{zeta2}) and (\ref{R_2_i}), neglecting $\delta \epsilon^{(1)}$ and $V_i$, and making use of $\z^{(1)} + \R^{(1)} \approx 0$, one can show that the spatial component of Eq.~(\ref{zeta_R_ls}) yields $\z^{(2)} + \R^{(2)} \approx 0$ \cite{Vernizzi:2004nc}.
From this relation and the second-order evolution equation of $\zeta$,
Eq.~(\ref{zeta2_evol}), one can find a large-scale evolution
equation for $\R$ at second order,
\beq
\R^{(2)}{}'\approx \frac{H}{\sib'^2} \left[ 2\bar \theta' \sib'
\ss
 -   \left( \bar V_{,ss} + 4 \bar \theta'{}^2
\right)  \sf^2  
+ \frac{\bar V_{,\sigma}}{\bar \sigma'} \delta s \delta s'
 \right]. \label{R2_evol}
\eeq
 The second-order perturbation $\R^{(2)}$ can be related on large scales to
$Q_{\rm SM}^{(2)}$ by combining Eqs.~(\ref{Q_2}) and  (\ref{R_2})
and using Eq.~(\ref{relation_added}). One obtains
\beq
\R^{(2)} \approx \frac{H}{\sib' } \left[Q_{\rm SM}^{(2)}
-\frac{1}{\sib'} \left(Q_{\rm SM}'- \thetab'\sf \right) Q_{\rm SM}
- \frac{1}{2H} \left(\frac{H}{\sib'} \right)' Q_{\rm SM}^2
\right], \label{RQ_rel}
\eeq
 which can be
used to show that Eq.~(\ref{R2_evol}) can be rewritten as a first integral for $\Qs_{\rm SM}$.

Let us now discuss the second order evolution of the entropy perturbation $\sf$. 
On  {\em large scales} this is obtained by  simply expanding  the
spatial components of Eq.~(\ref{s_evol_ls}) up to second order, which gives
 \beq
 \begin{split}
\ss{}''+3H \ss{}'+\left(\bar V_{,ss}+3{\bar {\theta}}^{\prime
2}\right)  \ss \approx   -\frac{\bar \theta'}{\sib'}  \sf'{}^2
- \frac{2}{\sib'}\left( \bar \theta''+ \bar \theta'
\frac{\bar V_{,\sigma}}{\sib'} -  \frac{3}{2} H \thetab'\right)
\sf \sf'
  - \left( \frac{1}{2} \bar V_{,sss} - 5\frac{\bar
\theta'}{\sib'} \bar V_{,ss} - 9 \frac{\bar \theta'{}^3}{\sib'}
\right)\sf^2  .
\label{s_evol_2}
\end{split}
 \eeq
As in the linear theory,  the entropy perturbation evolves
independently of the adiabatic component on large scales.

\subsubsection{Third-order perturbations}
\label{sec:thirdorder_fields}

\newcommand{\be}{\begin{equation}}
\newcommand{\ee}{\end{equation}}
\def\s{{\sigma}}
\def\sb{\bar{\sigma}}
\def\x{\xi}
\def\tb{\bar{\theta}}
\def\half{\frac{1}{2}}
\def\pt{\partial}
\def\th{\theta}
\def\tb{\bar{\theta}}

It is straightforward to extend the same procedure at third order, as discussed in detail  in \cite{Lehners:2009ja}. The evolution equation for $\z^{(3)}$ on large scales can be obtained by expanding up to third order Eq.~(\ref{zeta_evol_ls}), which yields an equation of the form
\beq
\z^{(3)}{}' \approx  {\cal S}_\zeta^{(3)} \, [\d s, \d s', \d s^{(2)}, \d s^{(2)\prime}],
\label{zeta3'}
\eeq
where the source term on the right-hand side contains  terms cubic in first order perturbations ($\d s$ or  $\d s'$) 
and terms that are products of second order perturbations ($\d s^{(2)}$ or $\d s^{(2)\prime}$) with first order perturbations ($\d s$ or  $\d s'$). 
The explicit definition of $\d s^{(3)}$ used in this expansion is 
\beq
\delta
s^{(3)}\equiv \bar{e}_{s I} \delta \phi^{I (3)} -\frac{\delta
\s^{(2)}}{\sb'}\left(\delta s'+\tb' \delta \s
\right)-\frac{\delta \s}{\sb'}\delta s^{(2)'}-\frac{\delta
\s^2}{2 \sb^{'2}}  (\d s'' - \frac{\sb''}{\sb} \d s' +\tb^{'2}
\delta s)  -\frac{\delta \s^3}{6 \sb'}\left(\frac{\tb'}{\sb'}
\right)' -\frac{\tb'}{2 \sb^{'2}}\delta s \delta s'
\delta \s, \label{s3} 
\eeq
 which can be determined by an appropriate expansion of  $s_a$ at third order, similar to the second order expansion 
 (\ref{s_i}).

To complete this equation, one needs to derive the
second order (in time) equation of motion at third order on large scales.
This is done by expanding Eq.~(\ref{s_evol_ls}) up to third order,
which yields an equation of the form
\beq
\label{eom_s_3}
 \d s^{(3)''} + 3H\d s^{(3)'} +
(\bar V_{,ss} + 3\tb^{'2}) \d s^{(3)} 
\approx {\cal S}_s^{(3)} \, [\d s, \d s', \d s^{(2)}, \d s^{(2)\prime}].
\eeq 
The explicit expressions of the source terms ${\cal S}_\z$ and ${\cal S}_s$, which  have been derived explicitly in \cite{Lehners:2009ja}, are given in the Appendix.

\subsection{An application: super-Hubble generation of non-Gaussian perturbations }
An important application of the above formalism is the computation of the non-Gaussianities 
of the primordial curvature perturbation. 
The computation of  $\zeta$ and $\delta s$, by integration of their respective evolution equations up to the required order, can be seen as an alternative to the $\delta N$ formalism, described in the article by Wands in this issue~\cite{wands}.
It is especially useful when the background evolution is complicated so that one cannot express analytically the number of e-folds as a function of the initial scalar field amplitudes.

Formally, one can write the solutions of the evolution equations as
\bea
\z & = & \z_* + T_{\z }^{(1)} \delta s_* + T_{\z }^{(2)}
\delta s_*^2+ T_{\z }^{(3)}
\delta s_*^3+\dots
\;, \label{gensolzeta} \\
  \delta s & = & T^{(1)}_{s} \delta s_* +
T^{(2)}_{s} \delta s_*^2+
T^{(3)}_{s} \delta s_*^3 +\dots\;, \label{gensols}
 \eea
where  the $T_{\zeta}^{(n)}$ and $T_{s}^{(n)}$ correspond to  transfer functions, for $\z$ and $\delta
s$, respectively, at the $n$-th order and $\z_*$, $\delta s_*$ are their initial
conditions at Hubble exit. In particular, we are interested in the values of $\z$ and $\d s$ after inflation, at the onset of the radiation dominated era.

Neglecting slow-roll corrections, the field perturbations $Q_{\rm SM *}$ and $\delta s_*$ at Hubble crossing can be treated as Gaussian, with respective power spectra  $P_{Q_*}(k)$ and $P_{\d s_*}(k)$ (see Appendix of \cite{Langlois:2008vk} for an explicit calculation of their 3-point functions). Note that there exist models with non-standard kinetic terms where the perturbations just after horizon crossing are  non-Gaussian (see \cite{Langlois:2010xc} and the article of K. Koyama in this issue~\cite{Koyama:2010xj} for a general discussion on multi-field models of this type) and where the entropy perturbations can play a crucial r\^ole~\cite{DBI}. However, we will not consider these models here. 

The perturbations  $Q_{\rm SM *}$ and $\delta s_*$ are also {\em independent} random fields, at least at leading order in slow-roll \cite{Byrnes:2006fr,Lalak:2007vi}. 
Moreover, although the  relation between $\zeta$ and $Q_{\rm SM}$ is in principle nonlinear (as illustrated for example by
Eq.~(\ref{RQ_rel}) for $\R$ which coincides with $\zeta$ on large scales), at horizon crossing the linear  relation dominates in slow-roll models so that 
$\z_* \simeq - (H/\sb') Q_{\rm SM *}$ \cite{Langlois:2008vk}.

\subsubsection{Curvature perturbations}

Using Eqs.~(\ref{gensolzeta}) and (\ref{gensols}) one finds that  the final 2-point, 3-point and 4-point functions  of $\z$, which are in principle  observable,  are given respectively by
\bea
\langle \z_{\vec k} \z_{\vec k'} \rangle & =& (2\pi)^3 \delta (\vec k+\vec k') P_\z (k), \qquad P_\z (k) \equiv  (2 \epsilon_* \mP^2)^{-1} P_{Q_*}(k) + (T_\z^{(1)})^2P_{\d s_*}(k)  , \label{PS_zeta}\\
\langle \z_{\vec k_1} \z_{\vec k_2} \z_{\vec k_3} \rangle &\simeq& 2 (T_\z^{(1)})^2 T_\z^{(2)} \left[ P_{\d s_*}(k_1) P_{\d s_*}(k_2) + 2 \ {\rm cyclic} \right], \label{3-point}\\
\langle \z_{\vec k_1} \z_{\vec k_2} \z_{\vec k_3} \z_{\vec k_4} \rangle & \simeq &  4 (T_\z^{(1)})^2 (T_\z^{(2)})^2 \left[ P_{\d s_*}(k_1) P_{\d s_*}(k_2) P_{\d s_*}(k_{13}) + 11 \ {\rm perms} \right] \nonumber \\
&& + 6 (T_\z^{(1)})^3 T_\z^{(3)} \left[ P_{\d s_*}(k_1) P_{\d s_*}(k_2)P_{\d s_*}(k_3) + 3 \ {\rm cyclic} \right], \label{4-point}
\eea
where we have introduced the slow-roll parameter $\epsilon \equiv  4\pi G \sb'^2/H^2$ and $M_P\equiv (8\pi G)^{-1/2}$.

If the power spectrum is dominated by its entropy  contribution, i.e. $ (T_\z^{(1)})^2P_{\d s_*}\gg (2 \epsilon_* \mP^2)^{-1} P_{Q_*}$, then one can easily relate the non-Gaussianities generated by the isocurvature field to the {\em local} nonlinear parameters $f_{\rm NL}$, $\tau_{\rm NL}$ and $g_{\rm NL}$, defined by \cite{Maldacena:2002vr,Byrnes:2006vq,Lyth:2005fi,Seery:2008ax}
\beq
\langle \z_{\vec k_1} \z_{\vec k_2} \z_{\vec k_3} \rangle= \frac65 f_{\rm NL} \left[ P_\z(k_1) P_\z(k_2) + 2 \ {\rm cyclic} \right],
\eeq
\beq
\langle \z_{\vec k_1} \z_{\vec k_2} \z_{\vec k_3} \z_{\vec k_4} \rangle  =   \tau_{\rm NL} \left[ P_\z(k_1) P_\z(k_2) P_\z(k_{13}) + 11 \ {\rm perms} \right] 
 + \frac{54}{25} g_{\rm NL} \left[ P_\z(k_1) P_\z(k_2)P_\z(k_3) + 3 \ {\rm cyclic} \right].
\eeq
From Eqs.~(\ref{3-point}) and (\ref{4-point}) one finds
\beq
f_{\rm NL} = \frac53 \frac{T^{(2)}_\z}{(T^{(1)}_\z)^2}, \qquad \tau_{\rm NL} = 4 \frac{(T^{(2)}_\z)^2}{(T^{(1)}_\z)^4}, \qquad g_{\rm NL} = \frac{25}{9} \frac{T^{(3)}_\z}{(T^{(1)}_\z)^3}. \label{nonlinear_par}
\eeq

These expressions can be applied to compute the non-Gaussianities generated in multifield models. 
For instance, in the case of two fields, for separable potentials -- potentials that can be written as the sum or product of two functions dependent only on one of the fields, the transfer functions can be computed analytically either from the evolution equations for adiabatic and isocurvature perturbations derived in this section or by using the $\delta N$-formalism \cite{Vernizzi:2006ve,Byrnes:2008wi,Seery:2006js}. It has been shown that the non-Gaussianity generated after inflation are slow-roll suppressed in these cases (see also the review by Byrnes and Choi \cite{Byrnes:2010em}). The non-linear evolution of the isocurvature field during inflation can generate a large non-vanishing 3 and 4-point functions \cite{Bernardeau:2002jy,Rigopoulos:2005ae}, as reviewed by Bernardeau in this issue \cite{bernardeau}. From Eq.~(\ref{nonlinear_par}), the conditions to obtain large nonlinear parameters are ${T^{(2)}_\z} \gg {(T^{(1)}_\z)^2}$ and $T^{(3)}_\z \gg {(T^{(1)}_\z)^3}$. 

Eqs.~(\ref{nonlinear_par}), together with the evolution equations for adiabatic and isocurvature perturbations derived in this section, have  also been applied to compute the non-Gaussianities generated during the new Ekpyrotic scenario \cite{Lehners:2007wc,Lehners:2008my,Lehners:2009ja,Lehners:2010fy} with results that agree with those obtained using the $\delta N$-formalism \cite{Koyama:2007if,Lehners:2009ja}.

A similar approach (although keeping the original scalar fields instead of a decomposition into adiabatic and entropic modes) has been used to compute the evolution of $\zeta$ up to second order, and thus obtain the non-Gaussianities, in the context of hybrid inflation~\cite{Enqvist:2004bk,Jokinen:2005by,Barnaby:2006cq,Barnaby:2006km}. In these works the curvature perturbation at second order was expressed as a time integral of an expression quadractic in the linear perturbations of the tachyonic field. The same procedure can also been used to study the preheating phase at the end of hybrid inflation~\cite{Barnaby:2006cq,Barnaby:2006km}.

\subsubsection{Isocurvature perturbations}

It is possible, although not necessary, that the entropy perturbations generated during inflation survive after inflation, in the form of traditional entropy or  isocurvature  perturbations, i.e.~fluctuations  of the relative number of particles between two species (for instance, the number of photons per cold dark matter particle).  
In  this case, the accessible cosmological data would result from a combination of curvature and  isocurvature perturbations, which could be correlated~\cite{Langlois:1999dw}. 
So far, there has been no detection of an isocurvature perturbation in the CMB data and the constraints on the power spectra are already tight~\cite{Komatsu:2010fb}.  One could also envisage that an isocurvature component might  be detectable through its own higher-order correlations or its higher-order correlations with the adiabatic component, as discussed in 
\cite{Langlois:2008vk,Kawasaki:2008sn,Kawasaki:2008pa,Kawakami:2009iu,Takahashi:2009cx}. Such a possibility can arise for instance in the mixed curvaton and
inflaton scenario introduced in \cite{Langlois:2004nn}.

In the radiation dominated era the adiabatic perturbation coincides with the perturbation in the radiation fluid which, using Eq.~(\ref{zeta_a_w}) with $w=1/3$,  can be written as 
\beq
\z_a^{\rm r} =  \partial_a \zeta, \qquad \zeta  \equiv \delta \alpha + \frac14  \ln \left( \frac{\rho_{\rm r}}{\rb_{\rm r}}  \right) .
\eeq
The isocurvature perturbation between the cold dark matter and radiation, say, can be defined as 
\beq
S_a =  3 ( \z_a^{\rm c} - \z_a^{\rm r}) = \partial_a S, \qquad S \equiv \ln \left( \frac{\rho_{\rm c}}{\rb_{\rm c}}  \right) -  \frac34  \ln \left( \frac{\rho_{\rm r}}{\rb_{\rm r}}  \right) .
\eeq
Formally, one can write the relation between the entropy perturbation during inflation and the final isocurvature perturbation in the form
 \beq
  S = T^{(1)}_S \delta s_* + T^{(2)}_S \delta s_*^2+
T^{(3)}_S \delta s_*^3 +\dots\,  \;, \label{gensolS}
 \eeq
 where the transfer functions $T_S^{(n)}$ depend on the details of the model. Analytically solvable examples of $T_S^{(2)}$ are given for instance in \cite{Langlois:2008vk}.

The  observational upper limit on the  fraction of isocurvature fluctuations allowed by data  (see \cite{Komatsu:2010fb}) implies that the linear transfer coefficient $T_S^{(1)}$ must be very small. Interestingly, however,  $T_S^{(2)}$ and $T_S^{(3)}$ could still be large and the non-adiabatic perturbation detectable through its effect on higher-order correlation functions.  Indeed, neglecting   $T_S^{(1)}$, one finds that the non-vanishing 3 and 4-point correlation functions are given by 
\beq
\langle \z_{\vec k_1} \z_{\vec k_2} S_{\vec k_3} \rangle \simeq  2 (T_\z^{(1)})^2 T_S^{(2)}  P_{\d s_*}(k_1) P_{\d s_*}(k_2), 
\eeq
\beq
\langle \z_{\vec k_1} \z_{\vec k_2} S_{\vec k_3} S_{\vec k_4} \rangle \simeq  4 (T_\z^{(1)})^2 (T_S^{(2)})^2  P_{\d s_*}(k_1) P_{\d s_*}(k_2) \left[ P_{\d s_*}(k_{13})+ P_{\d s_*}(k_{23})\right], 
\eeq
\beq
\begin{split}
\langle \z_{\vec k_1} \z_{\vec k_2} \z_{\vec k_3} S_{\vec k_4} \rangle & \simeq  4 (T_\z^{(1)})^2 T_\z^{(2)} T_S^{(2)}  \{ P_{\d s_*}(k_1) P_{\d s_*}(k_2) \left[  P_{\d s_*}(k_{13}) +  P_{\d s_*}(k_{23})  \right] + 2 \ {\rm cyclic} (1,2,3) \} \\ & + 6 (T_\z^{(1)})^3 T_S^{(3)} P_{\d s_*}(k_1) P_{\d s_*}(k_2) P_{\d s_*}(k_3). 
\end{split}
\eeq
It would thus be interesting  to obtain some observational constraints on the isocurvature non-Gaussianities  from the future data, such as those collected by  the Planck satellite.

\appendix

\section*{Appendix: source terms at third order}
The source terms that appear on the right hand side of (\ref{zeta3'}) 
and (\ref{eom_s_3}) are respectively given by~\cite{Lehners:2009ja}
\beq
\begin{split}
{\cal S}_\zeta^{(3)} \, [\d s, \d s', \d s^{(2)}, \d s^{(2)\prime}]  \equiv - &\frac{H}{\sb^{'2}}
\left[2 \tb' \sb' \d s^{(3)}  - 2( \bar V_{,ss}  + 4 \tb'{}^2)\d s \d s^{(2)}  + \frac{\bar V_{,\s}}{\sb'}(\d s^{(2)} \d s' )' \right.\\
+& \left.\left( \frac{\bar V_{,s\s}}{\sb'} - \frac{11}{3} \frac{\tb' \bar V_{,\s}}{\sb'{}^2} \right)\d s^2\d
s' + \left( 8 \frac{\tb'{}^3}{\sb'}+ 4 \frac{\tb' }{\sb'} \bar V_{,ss} - \frac{1}{3}\bar V_{,sss} \right)\d s^3\right] .
\end{split}
\eeq
and
\beq
\begin{split} 
& {\cal S}_s^{(3)} \, [\d s, \d s', \d s^{(2)}, \d s^{(2)\prime}]= - 2 \frac{\bar \theta'}{\sb'} \d s' \d s^{(2)}{}' \\  &-   
\frac{2}{\sb'} \left( \tb''+ \tb' \frac{\bar
V_{,\s}}{\sb'}-\frac32 H \tb' \right)  (\d s \d s^{(2)})'
 - \left( \bar V_{,sss}-10 \tb' \frac{ \bar
V_{,ss}}{\sb'}-18\frac{\tb^{'3}}{\sb'}\right)\d s \d s^{(2)}
 \\ & - \frac{\bar V_{,\s}}{\sb^{'3}}\d s'^3
 - \left(\frac{\bar V_{,\s\s}}{\sb^{'2}} +3\frac{\bar
V_{,\s}^2}{\sb^{'4}}+3H\frac{\bar
V_{,\s}}{\sb^{'3}}-2\frac{\bar
V_{,ss}}{\sb^{'2}}-6\frac{\tb^{'2}}{\sb^{'2}}\right) \d s'^2 \d
s  \\ &
+\left(10\frac{\tb'\tb''}{\sb^{'2}}+\frac{3}{2\sb'}\bar
V_{,ss\s}+5\frac{\bar V_{,\s}\bar V_{,ss}}{\sb^{'3}}
+7\frac{\tb^{'2}\bar V_{,\s}}{\sb^{'3}}+3H\frac{\bar
V_{,ss}}{\sb^{'2}}-14H\frac{\tb^{'2}}{\sb^{'2}}\right)\d s' \d
s^2  \\ & -\left(\frac{1}{6}\bar
V_{,ssss}-\frac{7}{3}\frac{\tb'}{\sb'}\bar V_{,sss}+2\frac{\bar
V_{,ss}^2}{\sb^{'2}} +21\frac{\tb^{'2}\bar
V_{,ss}}{\sb^{'2}}+27\frac{\tb^{'4}}{\sb^{'2}}\right)\d s^3  \,. \label{s3'} 
\end{split}
\eeq



\begin{thebibliography}{}

\bibitem{Bardeen:1980kt}
  J.~M.~Bardeen,
  Phys.\ Rev.\  D {\bf 22}, 1882 (1980).
  
\bibitem{Kodama:1985bj}
  H.~Kodama and M.~Sasaki,
  Prog.\ Theor.\ Phys.\ Suppl.\  {\bf 78}, 1 (1984).

\bibitem{Mukhanov:1990me}
  V.~F.~Mukhanov, H.~A.~Feldman and R.~H.~Brandenberger,
  Phys.\ Rept.\  {\bf 215}, 203 (1992).
  
\bibitem{Durrer:1993db}
  R.~Durrer,
  Fund.\ Cosmic Phys.\  {\bf 15}, 209 (1994)
  [arXiv:astro-ph/9311041].
  
\bibitem{Langlois:2004de}
  D.~Langlois, 
  ``Inflation, quantum fluctuations and cosmological perturbations'',  Cargese lectures 2003 [arXiv:hep-th/0405053].
  
\bibitem{Malik:2008im}
  K.~A.~Malik and D.~Wands,
  Phys.\ Rept.\  {\bf 475}, 1 (2009)
  [arXiv:0809.4944 [astro-ph]].

\bibitem{Langlois:2010xc}
  D.~Langlois,  ``Lectures on inflation and cosmological perturbations,''
  arXiv:1001.5259 [astro-ph.CO].

\bibitem{Tomita}
  K.~Tomita,
  Prog.\ Theor.\ Phys.\ {\bf 37}, 831 (1967). 

\bibitem{Bruni:1996im}
  M.~Bruni, S.~Matarrese, S.~Mollerach and S.~Sonego,
  Class.\ Quant.\ Grav.\  {\bf 14}, 2585 (1997)
  [arXiv:gr-qc/9609040].
  
\bibitem{Matarrese:1997ay}
  S.~Matarrese, S.~Mollerach and M.~Bruni,
  Phys.\ Rev.\  D {\bf 58}, 043504 (1998)
  [arXiv:astro-ph/9707278].

\bibitem{Bartolo:2004if}
  N.~Bartolo, E.~Komatsu, S.~Matarrese and A.~Riotto,
  Phys.\ Rept.\  {\bf 402}, 103 (2004)
  [arXiv:astro-ph/0406398].

\bibitem{Noh:2004bc}
  H.~Noh and J.~c.~Hwang,
  Phys.\ Rev.\  D {\bf 69}, 104011 (2004).

\bibitem{Salopek:1990jq}
  D.~S.~Salopek and J.~R.~Bond,
  Phys.\ Rev.\  D {\bf 42}, 3936 (1990).
  
\bibitem{Comer:1994np}
  G.~L.~Comer, N.~Deruelle, D.~Langlois and J.~Parry,
  Phys.\ Rev.\  D {\bf 49}, 2759 (1994).
  
\bibitem{Deruelle:1994iz}
  N.~Deruelle and D.~Langlois,
  Phys.\ Rev.\  D {\bf 52}, 2007 (1995)
  [arXiv:gr-qc/9411040].
  
\bibitem{Lyth:2003im}
  D.~H.~Lyth and D.~Wands,
  Phys.\ Rev.\  D {\bf 68}, 103515 (2003)
  [arXiv:astro-ph/0306498].
  
\bibitem{Rigopoulos:2003ak}
  G.~I.~Rigopoulos and E.~P.~S.~Shellard,
  Phys.\ Rev.\  D {\bf 68}, 123518 (2003)
  [arXiv:astro-ph/0306620].
    
\bibitem{Kolb:2004jg}
  E.~W.~Kolb, S.~Matarrese, A.~Notari and A.~Riotto,
  Mod.\ Phys.\ Lett.\  A {\bf 20}, 2705 (2005)
  [arXiv:astro-ph/0410541].
  
\bibitem{Lyth:2004gb}
  D.~H.~Lyth, K.~A.~Malik and M.~Sasaki,
  JCAP {\bf 0505}, 004 (2005)
  [arXiv:astro-ph/0411220].

\bibitem{Starobinsky:1986fxa}
  A.~A.~Starobinsky,
  JETP Lett.\  {\bf 42}, 152 (1985)
  [Pisma Zh.\ Eksp.\ Teor.\ Fiz.\  {\bf 42}, 124 (1985)].
  
\bibitem{Sasaki:1995aw}
  M.~Sasaki and E.~D.~Stewart,
  Prog.\ Theor.\ Phys.\  {\bf 95}, 71 (1996)
  [arXiv:astro-ph/9507001].
  
  \bibitem{Langlois:2005ii}
  D.~Langlois and F.~Vernizzi,
  Phys.\ Rev.\ Lett.\  {\bf 95}, 091303 (2005)
  [arXiv:astro-ph/0503416].

\bibitem{Langlois:2005qp}
  D.~Langlois and F.~Vernizzi,
  Phys.\ Rev.\ D {\bf 72}, 103501 (2005)
  [arXiv:astro-ph/0509078].

\bibitem{Langlois:2006iq}
  D.~Langlois and F.~Vernizzi,
  JCAP {\bf 0602}, 014 (2006)
  [arXiv:astro-ph/0601271].

\bibitem{Langlois:2006vv}
  D.~Langlois and F.~Vernizzi,
  JCAP {\bf 0702}, 017 (2007)
  [arXiv:astro-ph/0610064].
  
\bibitem{Hawking:1966qi}
  S.~W.~Hawking,
  Astrophys.\ J.\  {\bf 145}, 544 (1966).

\bibitem{Ellis}
G.~F.~R.~Ellis, Relativistic Cosmology, in {\em General Relativity
and Cosmology}, proceedings of the XLVII Enrico Fermi Summer
School, edited by R.~K.~Sachs (Academic, New York, 1971).
  
\bibitem{Ellis:1989jt}
  G.~F.~R.~Ellis and M.~Bruni,
  Phys.\ Rev.\  D {\bf 40}, 1804 (1989).

\bibitem{Ellis:1989ju}
  G.~F.~R.~Ellis, J.~Hwang and M.~Bruni,
  Phys.\ Rev.\  D {\bf 40}, 1819 (1989).
  
\bibitem{Ellis:1990gi}
  G.~F.~R.~Ellis, M.~Bruni and J.~Hwang,
  Phys.\ Rev.\  D {\bf 42}, 1035 (1990).
  
\bibitem{Bruni:1991kb}
  M.~Bruni, G.~F.~R.~Ellis and P.~K.~S.~Dunsby,
  Class.\ Quant.\ Grav.\  {\bf 9}, 921 (1992).
  
\bibitem{Dunsby:1991xk}
  P.~K.~S.~Dunsby, M.~Bruni and G.~F.~R.~Ellis,
  Astrophys.\ J.\  {\bf 395}, 54 (1992).
  
\bibitem{Bruni:1992dg}
  M.~Bruni, P.~K.~S.~Dunsby and G.~F.~R.~Ellis,
  Astrophys.\ J.\  {\bf 395}, 34 (1992).
  
  \bibitem{maartens}
  R. Maartens, this issue.
  
\bibitem{Clarkson:2003af}
  C.~A.~Clarkson,
  Phys.\ Rev.\  D {\bf 70}, 103524 (2004)
  [Erratum-ibid.\  D {\bf 70}, 129902 (2004)]
  [arXiv:astro-ph/0311505].

\bibitem{Bardeen:1983qw}
  J.~M.~Bardeen, P.~J.~Steinhardt and M.~S.~Turner,
  Phys.\ Rev.\  D {\bf 28}, 679 (1983).

\bibitem{Hwang}
  J.~C.~Hwang,
  Astrophys.\ J.\  {\bf 380}, 307 (1991).
    
  \bibitem{Wands:2000dp}
  D.~Wands, K.~A.~Malik, D.~H.~Lyth and A.~R.~Liddle,
  Phys.\ Rev.\  D {\bf 62}, 043527 (2000)
  [arXiv:astro-ph/0003278].
    
\bibitem{Wald:1984rg}
  R.~M.~Wald,
{\it  Chicago, Usa: Univ. Pr. (1984) 491p}

\bibitem{Malik:2003mv}
  K.~A.~Malik and D.~Wands,
  Class.\ Quant.\ Grav.\  {\bf 21}, L65 (2004)
  [arXiv:astro-ph/0307055].

\bibitem{Lyth:2005du}
  D.~H.~Lyth and Y.~Rodriguez,
  Phys.\ Rev.\  D {\bf 71}, 123508 (2005)
  [arXiv:astro-ph/0502578].
  
\bibitem{Enqvist:2006fs}
  K.~Enqvist, J.~Hogdahl, S.~Nurmi and F.~Vernizzi,
  Phys.\ Rev.\  D {\bf 75}, 023515 (2007)
  [arXiv:gr-qc/0611020].
  
\bibitem{Lehners:2009ja}
  J.~L.~Lehners and S.~Renaux-Petel,
  Trispectrum,''
  Phys.\ Rev.\  D {\bf 80}, 063503 (2009)
  [arXiv:0906.0530 [hep-th]].
  
\bibitem{Langlois:2008mn}
  D.~Langlois and S.~Renaux-Petel,
  JCAP {\bf 0804}, 017 (2008)
  [arXiv:0801.1085 [hep-th]].

\bibitem{RenauxPetel:2008gi}
  S.~Renaux-Petel and G.~Tasinato,
  JCAP {\bf 0901}, 012 (2009)
  [arXiv:0810.2405 [hep-th]].
  
\bibitem{Gordon:2000hv}
  C.~Gordon, D.~Wands, B.~A.~Bassett and R.~Maartens,
  Phys.\ Rev.\  D {\bf 63}, 023506 (2001)
  [arXiv:astro-ph/0009131].

\bibitem{GrootNibbelink:2001qt}
  S.~Groot Nibbelink and B.~J.~W.~van Tent,
  Class.\ Quant.\ Grav.\  {\bf 19}, 613 (2002)
  [arXiv:hep-ph/0107272].
  
\bibitem{Rigopoulos:2005xx}
  G.~I.~Rigopoulos, E.~P.~S.~Shellard and B.~J.~W.~van Tent,
  Phys.\ Rev.\  D {\bf 73}, 083521 (2006)
  [arXiv:astro-ph/0504508].

\bibitem{Pitrou:2007xy}
  C.~Pitrou and J.~P.~Uzan,
  Phys.\ Rev.\  D {\bf 75}, 087302 (2007)
  [arXiv:gr-qc/0701121].
      
\bibitem{Sasaki:1986hm}
  M.~Sasaki,
  Prog.\ Theor.\ Phys.\  {\bf 76}, 1036 (1986).
  
\bibitem{Mukhanov:1988jd}
  V.~F.~Mukhanov,
  Sov.\ Phys.\ JETP {\bf 67}, 1297 (1988)
  [Zh.\ Eksp.\ Teor.\ Fiz.\  {\bf 94N7}, 1 (1988)].

\bibitem{Taruya:1997iv}
  A.~Taruya and Y.~Nambu,
  Phys.\ Lett.\  B {\bf 428}, 37 (1998)
  [arXiv:gr-qc/9709035].
  
\bibitem{Acquaviva:2002ud}
  V.~Acquaviva, N.~Bartolo, S.~Matarrese and A.~Riotto,
  Nucl.\ Phys.\  B {\bf 667}, 119 (2003)
  [arXiv:astro-ph/0209156].

\bibitem{Vernizzi:2004nc}
  F.~Vernizzi,
  Phys.\ Rev.\  D {\bf 71} (2005) 061301
  [arXiv:astro-ph/0411463].
  
\bibitem{Malik:2005cy}
  K.~A.~Malik,
  JCAP {\bf 0511}, 005 (2005)
  [arXiv:astro-ph/0506532].

\bibitem{Malik:2006ir}
  K.~A.~Malik,
  JCAP {\bf 0703}, 004 (2007)
  [arXiv:astro-ph/0610864].
  
\bibitem{Lyth:2005fi}
  D.~H.~Lyth and Y.~Rodriguez,
  Phys.\ Rev.\ Lett.\  {\bf 95}, 121302 (2005)
  [arXiv:astro-ph/0504045].

\bibitem{Vernizzi:2006ve}
  F.~Vernizzi and D.~Wands,
  JCAP {\bf 0605}, 019 (2006)
  [arXiv:astro-ph/0603799].

\bibitem{Maldacena:2002vr}
  J.~Maldacena,
  JHEP {\bf 0305}, 013 (2003)
  [arXiv:astro-ph/0210603].
  
  \bibitem{wands} D. Wands, this issue.
  
\bibitem{Langlois:2008vk}
  D.~Langlois, F.~Vernizzi and D.~Wands,
  JCAP {\bf 0812}, 004 (2008)
  [arXiv:0809.4646 [astro-ph]].

\bibitem{Koyama:2010xj}
  K.~Koyama,
  arXiv:1002.0600 [hep-th].
  
  \bibitem{DBI}
 D.~Langlois, S.~Renaux-Petel, D.~A.~Steer and T.~Tanaka,
 Phys.\ Rev.\ Lett.\  {\bf 101},  061301 (2008)
  [arXiv:0804.3139 [hep-th]]

\bibitem{Byrnes:2006fr}
  C.~T.~Byrnes and D.~Wands,
  Phys.\ Rev.\  D {\bf 74}, 043529 (2006)
  [arXiv:astro-ph/0605679].

\bibitem{Lalak:2007vi}
  Z.~Lalak, D.~Langlois, S.~Pokorski and K.~Turzynski,
  JCAP {\bf 0707}, 014 (2007)
  [arXiv:0704.0212 [hep-th]].

\bibitem{Byrnes:2006vq}
  C.~T.~Byrnes, M.~Sasaki and D.~Wands,
  Phys.\ Rev.\  D {\bf 74}, 123519 (2006)
  [arXiv:astro-ph/0611075].

\bibitem{Seery:2008ax}
  D.~Seery, M.~S.~Sloth and F.~Vernizzi,
  JCAP {\bf 0903}, 018 (2009)
  [arXiv:0811.3934 [astro-ph]].


\bibitem{Byrnes:2008wi}
  C.~T.~Byrnes, K.~Y.~Choi and L.~M.~H.~Hall,
  JCAP {\bf 0810}, 008 (2008)
  [arXiv:0807.1101 [astro-ph]].

\bibitem{Seery:2006js}
  D.~Seery and J.~E.~Lidsey,
  JCAP {\bf 0701}, 008 (2007)
  [arXiv:astro-ph/0611034].

\bibitem{Byrnes:2010em}
  C.~T.~Byrnes and K.~Y.~Choi,
  arXiv:1002.3110 [astro-ph.CO].

\bibitem{Bernardeau:2002jy}
  F.~Bernardeau and J.~P.~Uzan,
  Phys.\ Rev.\  D {\bf 66}, 103506 (2002)
  [arXiv:hep-ph/0207295].

\bibitem{Rigopoulos:2005ae}
  G.~I.~Rigopoulos, E.~P.~S.~Shellard and B.~J.~W.~van Tent,
  Phys.\ Rev.\  D {\bf 73}, 083522 (2006)
  [arXiv:astro-ph/0506704].

\bibitem{bernardeau} F. Bernardeau, this issue.

\bibitem{Lehners:2007wc}
  J.~L.~Lehners and P.~J.~Steinhardt,
  Phys.\ Rev.\  D {\bf 77}, 063533 (2008)
  [Erratum-ibid.\  D {\bf 79}, 129903 (2009)]
  [arXiv:0712.3779 [hep-th]].

\bibitem{Lehners:2008my}
  J.~L.~Lehners and P.~J.~Steinhardt,
  Phys.\ Rev.\  D {\bf 78}, 023506 (2008)
  [Erratum-ibid.\  D {\bf 79}, 129902 (2009)]
  [arXiv:0804.1293 [hep-th]].

\bibitem{Lehners:2010fy}
  J.~L.~Lehners,
  arXiv:1001.3125 [hep-th].

\bibitem{Koyama:2007if}
  K.~Koyama, S.~Mizuno, F.~Vernizzi and D.~Wands,
  JCAP {\bf 0711}, 024 (2007)
  [arXiv:0708.4321 [hep-th]].
  
    \bibitem{Enqvist:2004bk}
  K.~Enqvist and A.~Vaihkonen,
  ``Non-Gaussian perturbations in hybrid inflation,''
  JCAP {\bf 0409}, 006 (2004)
  [arXiv:hep-ph/0405103].
  
\bibitem{Jokinen:2005by}
  A.~Jokinen and A.~Mazumdar,
  JCAP {\bf 0604}, 003 (2006)
  [arXiv:astro-ph/0512368].

\bibitem{Barnaby:2006cq}
  N.~Barnaby and J.~M.~Cline,
  Phys.\ Rev.\  D {\bf 73}, 106012 (2006)
  [arXiv:astro-ph/0601481].
  
\bibitem{Barnaby:2006km}
  N.~Barnaby and J.~M.~Cline,
  Phys.\ Rev.\  D {\bf 75}, 086004 (2007)
  [arXiv:astro-ph/0611750].

\bibitem{Langlois:1999dw}
  D.~Langlois,
  Phys.\ Rev.\  D {\bf 59}, 123512 (1999)
  [arXiv:astro-ph/9906080].
  
\bibitem{Komatsu:2010fb}
  E.~Komatsu {\it et al.},
  arXiv:1001.4538 [astro-ph.CO].

\bibitem{Kawasaki:2008sn}
  M.~Kawasaki, K.~Nakayama, T.~Sekiguchi, T.~Suyama and F.~Takahashi,
  JCAP {\bf 0811}, 019 (2008)
  [arXiv:0808.0009 [astro-ph]].
  
\bibitem{Kawasaki:2008pa}
  M.~Kawasaki, K.~Nakayama, T.~Sekiguchi, T.~Suyama and F.~Takahashi,
  JCAP {\bf 0901}, 042 (2009)
  [arXiv:0810.0208 [astro-ph]].
  
\bibitem{Kawakami:2009iu}
  E.~Kawakami, M.~Kawasaki, K.~Nakayama and F.~Takahashi,
  JCAP {\bf 0909}, 002 (2009)
  [arXiv:0905.1552 [astro-ph.CO]].

\bibitem{Takahashi:2009cx}
  T.~Takahashi, M.~Yamaguchi and S.~Yokoyama,
  Phys.\ Rev.\  D {\bf 80}, 063524 (2009)
  [arXiv:0907.3052 [astro-ph.CO]].

\bibitem{Langlois:2004nn}
  D.~Langlois and F.~Vernizzi,
  Phys.\ Rev.\  D {\bf 70}, 063522 (2004)
  [arXiv:astro-ph/0403258].
 
\end{thebibliography}
\end{document}